\documentclass[a4paper,12pt,dvips,oneside]{article}
\usepackage{booktabs,graphicx,overcite,xspace,amsmath,amssymb,longtable,multirow}
\usepackage[scanall]{psfrag}
\usepackage[small,sc]{caption}
\usepackage[dvips]{color}

\begin{document}
\def\aciee{Angew. Chem. Int. Ed. Engl. }
\def\ac{Acta. Crystallogr. }
\def\acp{Adv. Chem. Phys. }
\def\acr{Acc. Chem. Res. }
\def\ajp{Am. J. Phys. }
\def\ap{Ann. Physik }
\def\apc{Adv. Prot. Chem. }
\def\arpc{Ann. Rev. Phys. Chem. }
\def\cccc{Coll. Czech. Chem. Comm. }
\def\cpc{Comp. Phys. Comm. }
\def\cpl{Chem. Phys. Lett. }
\def\crev{Chem. Rev. }
\def\el{Europhys. Lett. }
\def\ic{Inorg. Chem. }
\def\ijmpc{Int. J. Mod. Phys. C }
\def\ijqc{Int. J. Quant. Chem. }
\def\jcis{J. Colloid Interface Sci. }
\def\jcsft{J. Chem. Soc., Faraday Trans. }
\def\jacs{J. Am. Chem. Soc. }
\def\jas{J. Atmos. Sci. }
\def\jbc{J. Biol. Chem. }
\def\jcc{J. Comp. Chem. }
\def\jcp{J. Chem. Phys. }
\def\jce{J. Chem. Ed. }
\def\jcscc{J. Chem. Soc., Chem. Commun. }
\def\jetp{J. Exp. Theor. Phys. (Russia) }
\def\jmb{J. Mol. Biol. }
\def\jmsp{J. Mol. Spec. }
\def\jmst{J. Mol. Struct. }
\def\jncs{J. Non-Cryst. Solids }
\def\jpa{J. Phys. A }
\def\jpc{J. Phys. Chem. }
\def\jpca{J. Phys. Chem. A }
\def\jpcb{J. Phys. Chem. B }
\def\jpcm{J. Phys. Condensed Matter. }
\def\jpcs{J. Phys. Chem. Solids. }
\def\jpsj{J. Phys. Soc. Jpn. }
\def\jrnist{J. Res. Natl. Inst. Stand. Technol. }
\def\mg{Math. Gazette }
\def\mp{Mol. Phys. }
\def\nat{Nature }
\def\nsb{Nat. Struct. Biol.}
\def\Pa{Physica A }
\def\pac{Pure. Appl. Chem. }
\def\pccp{Phys. Chem. Chem. Phys. }
\def\phys{Physics }
\def\pmb{Philos. Mag. B }
\def\ptrsb{Philos. T. Roy. Soc. B }
\def\pnasu{Proc. Natl. Acad. Sci. USA }
\def\pr{Phys. Rev. }
\def\prep{Phys. Reports }
\def\pra{Phys. Rev. A }
\def\prb{Phys. Rev. B }
\def\prbcm{Phys. Rev. B }
\def\prc{Phys. Rev. C }
\def\prd{Phys. Rev. D }
\def\pre{Phys. Rev. E }
\def\prl{Phys. Rev. Lett. }
\def\prsa{Proc. R. Soc. A }
\def\psfg{Proteins: Struct., Func. and Gen. }
\def\sci{Science }
\def\spj{Sov. Phys. JETP }
\def\ss{Surf. Sci. }
\def\tca{Theor. Chim. Acta }
\def\zpb{Z. Phys. B. }
\def\zpc{Z. Phys. Chem. }
\def\zpd{Z. Phys. D }
\def\aciee{Angew. Chem. Int. Ed. Engl. }
\def\ac{Acta. Crystallogr. }
\def\acp{Adv. Chem. Phys. }
\def\acr{Acc. Chem. Res. }
\def\ajp{Am. J. Phys. }
\def\ap{Adv. Phys. }
\def\arpc{Ann. Rev. Phys. Chem. }
\def\cccc{Coll. Czech. Chem. Comm. }
\def\cpl{Chem. Phys. Lett. }
\def\crev{Chem. Rev. }
\def\dalton{J. Chem. Soc., Dalton Trans. }
\def\el{Europhys. Lett. }
\def\faraday{J. Chem. Soc., Faraday Trans. }
\def\fartrans{J. Chem. Soc., Faraday Trans. }
\def\fdisc{J. Chem. Soc., Faraday Discuss. }
\def\ic{Inorg. Chem. }
\def\ijqc{Int. J. Quant. Chem. }
\def\jcis{J. Colloid Interface Sci. }
\def\jcsft{J. Chem. Soc., Faraday Trans. }
\def\jacs{J. Am. Chem. Soc. }
\def\jas{J. Atmos. Sci. }
\def\jcc{J. Comp. Chem. }
\def\jcp{J. Chem. Phys. }
\def\jce{J. Chem. Ed. }
\def\jcscc{J. Chem. Soc., Chem. Commun. }
\def\jetp{J. Exp. Theor. Phys. (Russia) }
\def\jmc{J. Math. Chem. }
\def\jmsp{J. Mol. Spec. }
\def\jmst{J. Mol. Structure }
\def\jncs{J. Non-Cryst. Solids }
\def\jpc{J. Phys. Chem. }
\def\jpcm{J. Phys. Condensed Matter. }
\def\jpsj{J. Phys. Soc. Jpn. }
\def\jsp{J. Stat. Phys. }
\def\mg{Math. Gazette }
\def\mp{Mol. Phys. }
\def\molphys{Mol. Phys. }
\def\nat{Nature }
\def\pac{Pure. Appl. Chem. }
\def\phys{Physics }
\def\pla{Phys. Lett. A }
\def\plb{Phys. Lett. B }
\def\phm{Philos. Mag. }
\def\pmb{Philos. Mag. B }
\def\pnas{Proc.\ Natl.\ Acad.\ Sci.\  USA }
\def\pr{Phys. Rev. }
\def\pra{Phys. Rev. A }
\def\prb{Phys. Rev. B }
\def\prc{Phys. Rev. C }
\def\prd{Phys. Rev. D }
\def\pre{Phys. Rev. E }
\def\prl{Phys. Rev. Lett. }
\def\prsa{Proc. R. Soc. A }
\def\ss{Surf. Sci. }
\def\sci{Science }
\def\tca{Theor. Chim. Acta }
\def\zpc{Z. Phys. Chem. }
\def\zpd{Z. Phys. D }
\def\zfpd{Z. Phys. D }
\def\zpdamc{Z. Phys. D }
\def\aciee{Angew. Chem. Int. Ed. Engl. }
\def\ac{Acta. Crystallogr. }
\def\acp{Adv. Chem. Phys. }
\def\acr{Acc. Chem. Res. }
\def\ajp{Am. J. Phys. }
\def\am{Adv. Mater. }
\def\apl{Appl. Phys. Lett. }
\def\arpc{Ann. Rev. Phys. Chem. }
\def\mrsb{Mater. Res. Soc. Bull. }
\def\cccc{Coll. Czech. Chem. Comm. }
\def\cj{Comput. J. }
\def\cp{Chem. Phys. }
\def\cpc{Comp. Phys. Comm. }
\def\cpl{Chem. Phys. Lett. }
\def\crev{Chem. Rev. }
\def\el{Europhys. Lett. }
\def\fd{Faraday Disc. }
\def\ic{Inorg. Chem. }
\def\ijmpc{Int. J. Mod. Phys. C }
\def\ijqc{Int. J. Quant. Chem. }
\def\jcis{J. Colloid Interface Sci. }
\def\jcsft{J. Chem. Soc., Faraday Trans. }
\def\jacs{J. Am. Chem. Soc. }
\def\jap{J. Appl. Phys. }
\def\jas{J. Atmos. Sci. }
\def\jcc{J. Comp. Chem. }
\def\jcp{J. Chem. Phys. }
\def\jce{J. Chem. Ed. }
\def\jcscc{J. Chem. Soc., Chem. Commun. }
\def\jetp{J. Exp. Theor. Phys. (Russia) }
\def\jmsp{J. Mol. Spec. }
\def\jmst{J. Mol. Structure }
\def\jncs{J. Non-Cryst. Solids }
\def\jpa{J. Phys. A }
\def\jpc{J. Phys. Chem. }
\def\jpcssp{J. Phys. C: Solid State Phys. }
\def\jpca{J. Phys. Chem. A. }
\def\jpcb{J. Phys. Chem. B. }
\def\jpcm{J. Phys. Condensed Matter. }
\def\jpcs{J. Phys. Chem. Solids. }
\def\jpsj{J. Phys. Soc. Jpn. }
\def\jpfmp{J. Phys. F, Metal Phys. }
\def\mg{Math. Gazette }
\def\mp{Mol. Phys. }
\def\msr{Mater. Sci. Rep. }
\def\nat{Nature }
\def\njc{New J. Chem. }
\def\pac{Pure. Appl. Chem. }
\def\phys{Physics }
\def\pma{Philos. Mag. A }
\def\pmb{Philos. Mag. B }
\def\pml{Philos. Mag. Lett. }
\def\pnasu{Proc. Natl. Acad. Sci. USA }
\def\pr{Phys. Rev. }
\def\prep{Phys. Reports }
\def\pra{Phys. Rev. A }
\def\prb{Phys. Rev. B }
\def\prc{Phys. Rev. C }
\def\prd{Phys. Rev. D }
\def\pre{Phys. Rev. E }
\def\prl{Phys. Rev. Lett. }
\def\prsa{Proc. R. Soc. A }
\def\pss{Phys. State Solidi }
\def\pssb{Phys. State Solidi B }
\def\rmp{Rev. Mod. Phys. }
\def\rpp{Rep. Prog. Phys. }
\def\sci{Science }
\def\ss{Surf. Sci. }
\def\tca{Theor. Chim. Acta }
\def\tetra{Tetrahedron }
\def\zpb{Z. Phys. B. }
\def\zpc{Z. Phys. Chem. }
\def\zpd{Z. Phys. D }

\title{Finding pathways between distant local minima}
\author{Joanne M. Carr, Semen A. Trygubenko
and David J. Wales\footnote{email: dw34@cam.ac.uk} \\
{\it University Chemical Laboratories, Lensfield Road,} \\
{\it Cambridge CB2 1EW, UK} }

\maketitle
\begin{abstract}
We report a new algorithm for constructing pathways between local minima that
involve a large number of intervening transition states on the potential energy
surface. A significant improvement in efficiency has been achieved by changing
the strategy for choosing successive pairs of local minima that serve as
endpoints for the next search. We employ Dijkstra's algorithm~\cite{dijkstra59} to identify the
`shortest' path corresponding to missing connections within an evolving database
of local minima and the transition states that connect them. The metric employed
to determine the shortest missing connection is a function of the minimised
Euclidean distance. We present applications to the formation of
buckminsterfullerene and to the folding of the B1 domain of protein G,
tryptophan zippers, and the villin headpiece subdomain. The corresponding pathways contain up to 163
transition states, and will be used in future discrete path sampling calculations.
\end{abstract}

\section{Introduction}

The discrete path sampling (DPS) approach \cite{Wales02,Wales03,Wales04} 
provides a means to sample pathways
corresponding to `rare events' using a coarse-grained framework based on the
underlying potential energy surface (PES). 
The PES can be formally partitioned into the catchment basins of local minima \cite{mezey81b},
while rate constants for transitions between these basins can be estimated using
statistical rate theory \cite{PelzerW32,Eyring35,EvansP35,forst73,Laidler87}
once transition states on the basin boundaries have been found.
Here we adopt the geometrical definition of
a transition state as a stationary point on the PES with a single negative Hessian eigenvalue, following
Murrell and Laidler.\cite{murrelll68} Minima are stationary points with no negative Hessian eigenvalues.

Two minima may be connected via a single transition
state if they are sufficiently close to each other in configuration space.
More generally, local minima may interconvert via any number of multi-step
paths, and the minimum number of elementary steps will depend upon the isomers
in question.\cite{Wales03}
Here we consider an elementary step as a rearrangement that involves a single transition state.
Locating multi-step pathways on a complex PES can be a difficult task. Double-ended
methods~\cite{Pratt86,elberk87,berrydb88,czerminskie90,FischerK92,StachoB92,
ionovac93,StachoB93,DomotorBS93,PengS93,matrofd94,smart94,BanDS94,MillsJ94,
ionovac95a,pengasf96,jonssonmj98,StachoDB99,ElberK99,
henkelmanj99,henkelmanuj00,henkelmanj00,HenkelmanJ01,MironF01,maragakisabrk02,PetersHBC04,trygubenkow04}
usually do not reveal all the
transition states at once, mainly because of the existence of multiple
barrier height and path length scales.\cite{trygubenkow04b}
Consecutive double-ended searches have proved to be an efficient way of working around
the above problem, and this strategy was adopted in our previous work.\cite{Wales02,trygubenkow04}
An essential part of this approach is a mechanism to incorporate the information
obtained in all the previous searches into the next one.
Various strategies can be adopted, the most general and effective being
the one based on Euclidean separation, and we refer the reader to the original publications for
details.\cite{Wales02,trygubenkow04}

A guess for the initial pathway is required for most double-ended searches.
Linear interpolation is a straightforward way to automate this part of the
calculation.\cite{henkelmanj99,henkelmanuj00,henkelmanj00,HenkelmanJ01,trygubenkow04}
For large endpoint separations guessing the initial pathway can be difficult,
and there is a large probability of finding many irrelevant stationary points
at the beginning of the calculation.

Due to the discrete nature of the double-ended methods such as
NEB~\cite{henkelmanuj00,henkelmanj00} and
DNEB~\cite{trygubenkow04}, which use a series of images to represent the path,
the candidate transition states obtained from such
searches usually need to be optimised further.\cite{maragakisabrk02,trygubenkow04}
For this purpose we employ
eigenvector-following methods.\cite{crippens71,pancir74,hilderbrandt77,%
cerjanm81,simonsjto83,banerjeeass85,baker86,wales94a,
walesu94,munrow97,munrow99,kumedamw01,walesw96}
The connectivity is obtained by following the two unique steepest-descent (SD) paths
downhill from each transition state. In our calculations the refinement of
transition states and
calculation of approximate steepest-descent paths are the most time consuming
steps.

The connection algorithm described in Ref.~\citen{trygubenkow04} uses
one double-ended search per cycle. However, we have found that this approach can
be overwhelmed by the abundance
of stationary points and pathways for complicated rearrangements.
In the present contribution we introduce the idea of an unconnected pathway and
make the connection algorithm more focused by allowing more than one
double-ended search per cycle.
This approach, in combination with some
other modifications described in the next section, greatly reduces the
computational demands of the method, and has allowed us to tackle more complicated problems. 
The new algorithm has been implemented within our {\tt OPTIM} package,
and a public domain version will be made available for download from the internet.\cite{download}
The examples presented below each involve the search for an initial
connection between distant endpoints for use in subsequent DPS calculations.

\section{Methods}
\label{sec:methods}

\subsection{Outline of the Connection Algorithm}
A detailed description of the previous version of our connection algorithm was provided
elsewhere~\cite{trygubenkow04}. Each iteration consists
of one double-ended transition state search, followed by refinement of
transition state candidates using
(hybrid) eigenvector-following~\cite{crippens71,pancir74,hilderbrandt77,cerjanm81,%
simonsjto83,banerjeeass85,baker86,wales94a,
walesu94,munrow97,munrow99,kumedamw01,walesw96}, and calculation
of approximate steepest-descent paths for each transition state to establish the connectivity.

We classify all the known local minima into
three categories: $S$ minima, which are connected to the starting structure, $F$
minima, which are connected to the final structure, and $U$ minima,
which are so far unconnected to the $S$ and $F$ sets.
$U$ minima can be connected between themselves but not to any $S$ or $F$
minima. When a connection is established between the members of $U$ and
$S$ or $F$ sets, the unconnected minimum, and all the minima connected to
it, become members of $S$ or $F$, respectively. If a connection is
found between members of the $S$ and $F$ sets then the algorithm terminates.

Before each cycle a decision must be made as to which minima
to try and connect next. Various strategies can be adopted, for example,
selection based on the order in which transition states were
found,\cite{Wales02}
or, selection of minima with the minimal separation in Euclidean distance
space.\cite{trygubenkow04}
However, when the endpoints are very distant in configuration space, neither of
these approaches is particularly efficient. The number of possible connections
that might be tried simply grows too quickly if the $S$, $F$ and $U$ sets become
large. However, the new algorithm described below seems to be very effective.

\subsection{A Dijkstra-based Selector}

The modified connection algorithm we have used in the present work
is based on a shortest path method proposed by Dijkstra~\cite{dijkstra59,cormenlrs01}.
We can describe the minima that are known at the beginning of each
connection cycle as a complete graph,\cite{CompleteGraph} $G=(M,E)$, where $M$ is the set of
all minima and $E$ is the set of all the edges between them.
Edges are considered to exist between every pair of minima $u$ and $v$, even if they are in
different $S$, $F$ or $U$ sets, and the weight of the edge is chosen to be a
function of the minimum Euclidean distance between them:\cite{rhee00}
\begin{equation}
w(u,v)=\left\{
\begin{array}{cl}
0, & \text{if $u$ and $v$ are connected via a single transition state,} \\
\infty, & \text{if~} n(u,v)=n_{max}, \\
f(D(u,v)), & \text{otherwise,}
\end{array}
\right.
\end{equation}
where $n(u,v)$ is the number of times a pair $(u,v)$ was selected for
a connection attempt, $n_{max}$ is the maximal number of times we may try to connect any
pair of minima, and $D(u,v)$ is the minimum Euclidean distance between $u$ and
$v$. $f$ should be a monotonically increasing function, such
as $f(D(u,v))=D(u,v)^2$.
We denote the number of minima in the set $M=S\cup U\cup F$, as $m$,
and the number of edges in the set $E$ as $e=m(m-1)/2$.

Using the Dijkstra algorithm~\cite{dijkstra59,cormenlrs01} and the weighted graph representation 
described above, it is possible to
determine the shortest paths between any minima in the database.
The source is selected to be one of the endpoints. Upon termination of the Dijkstra
algorithm, a shortest path from one endpoint to the other is extracted. If the
weight of this pathway is non-zero, it contains one or more `gaps'.
Connection attempts are then made for every pair $(u,v)$ of adjacent minima in the
pathway with non-zero $w(u,v)$ using the DNEB approach.\cite{trygubenkow04}

The computational complexity of the Dijkstra algorithm is at worst $O(m^2)$, and
the memory requirements scale in a similar fashion. The most
appropriate data structure is a weighted adjacency matrix. For the calculations
presented in this paper, the single source
shortest paths problem was solved at the beginning of each cycle, which took less than $10\%$
of the total execution time for the largest database encountered.
We emphasise here that once an initial path has been found, the perturbations
considered in typical DPS calculations will generally involve attempts to connect
minima that are separated by far fewer elementary rearrangements than the endpoints.
It is also noteworthy that the initial path is unlikely to be contribute significantly
to the overall rate constant. Nevertheless, it is essential to construct such a path
to begin the DPS procedure.

The nature of the definition of the weight function allows the Dijkstra algorithm 
to terminate whenever a second endpoint, or
any minimum connected to that endpoint via a series of elementary rearrangements, is reached.
This observation reduces the computational requirements by an amount that
depends on the distribution of the minima in the database among the $S$, $U$ and $F$ sets.
One of the endpoints is always a member of the $S$ set, while the other is a
member of $F$ set.
Either one can be chosen as the source, and we have found it most efficient to select
the one from the set with fewest members.
However, this choice does not improve the asymptotic bounds of the algorithm.

\section{Results}

\subsection{Buckminsterfullerene}

Various suggestions
\cite{HeathOCKS87,kroto88,Smalley92,mckaykw92,heath92,WakabayashiA92,
YeretzianHDW92,McElvanyRGD93,HunterFJ93b,WakabayashiSKA93,%
HeldenGB93,heldenhgb93,HunterFJ93,BabicT94,hunterj95,manolopoulosf96,BabicT96,strouts96,batess97}
have been made for the formation mechanism of buckminsterfullerene
\cite{krotohocs85,kratschmerlfh90} in the gas phase.
In total, there are 1812 different C$_{60}$ fullerene isomers \cite{liukss91,manolopoulosmd91},
which probably interconvert via the `pyracylene' or `SW' rearrangement \cite{stonew86}.
This process has been investigated in several previous studies
\cite{yib92,bakerf92,murrysos93,murryss94,HondaO96,walshw98,OsawaSH98,OsawaSHZ98},
and the most accurate calculations \cite{BettingerYS03,KumedaW03} 
yield a picture of the energy landscape 
that is quite similar to that obtained using a tight-binding potential \cite{walshw98,walesmw98}.
The same model \cite{widanyfkspjs96,porezagfsk95,seifertpf96}
was therefore used in this initial pathway calculation to minimise the 
computational expense, which is significantly greater than for analytical
empirical potentials \cite{marcoslra97}.
The present calculation therefore has most in common with
the annealing study of Xu and Scuseria \cite{xus94}, although the latter 
work did not involve transition state calculations.

\begin{figure}
\psfrag{V}[bl][bl]{$V$}
\psfrag{s}[bc][bc]{$s$}
\psfrag{-97.0}[cr][cr]{$-97.0$}
\psfrag{-97.2}[cr][cr]{$-97.2$}
\psfrag{-97.4}[cr][cr]{$-97.4$}
\psfrag{-97.6}[cr][cr]{$-97.6$}
\psfrag{-97.8}[cr][cr]{$-97.8$}
\psfrag{-98.0}[cr][cr]{$-98.0$}
\psfrag{-98.2}[cr][cr]{$-98.2$}
\psfrag{-98.4}[cr][cr]{$-98.4$}
\psfrag{-98.6}[cr][cr]{$-98.6$}
\psfrag{-98.3}[cr][cr]{$-98.3$}
\psfrag{-98.5}[cr][cr]{$-98.5$}
\psfrag{0}[tc][tc]{0}
\psfrag{200}[tc][tc]{200}
\psfrag{400}[tc][tc]{400}
\psfrag{600}[tc][tc]{600}
\psfrag{800}[tc][tc]{800}
\psfrag{1000}[tc][tc]{1000}
\psfrag{1200}[tc][tc]{1200}
\psfrag{1400}[tc][tc]{1400}
\psfrag{1600}[tc][tc]{1600}
\psfrag{1576}[tc][tc]{1576}
\psfrag{1580}[tc][tc]{1580}
\psfrag{1584}[tc][tc]{1584}
\psfrag{1588}[tc][tc]{1588}
\psfrag{1592}[tc][tc]{1592}
\psfrag{1596}[tc][tc]{1596}
\centerline{\includegraphics[height=0.95\textheight]{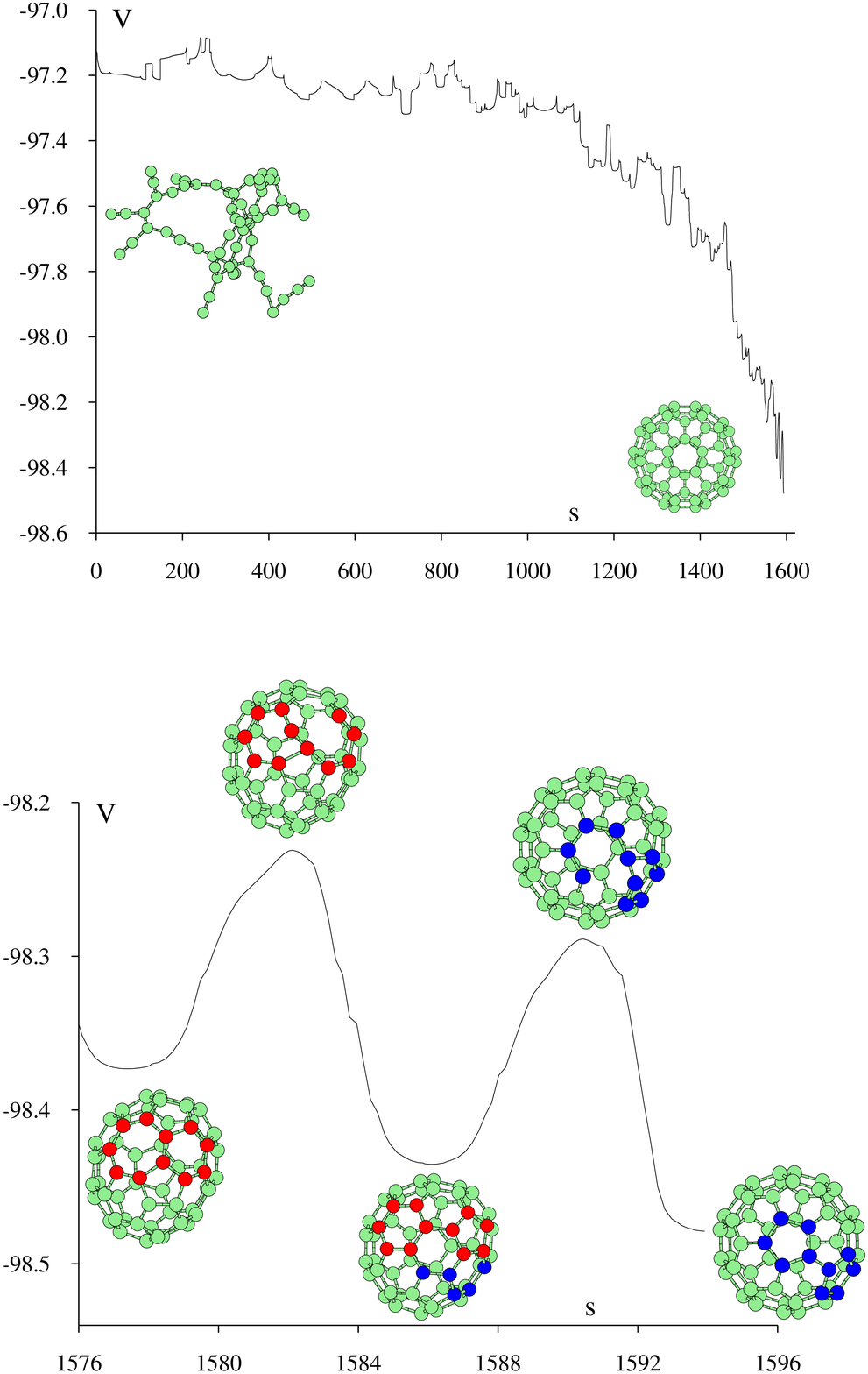}}
\ \vfill
\caption{}
\label{fig:C60}
\end{figure}

An initial high-energy starting point was constructed by simply placing sixty
carbon atoms in a container and minimising the energy. The resulting structure, which
contains a number of large rings and chains, including polyacetylene fragments,
is shown in Figure \ref{fig:C60}.
Although the structure of the buckminsterfullerene endpoint is well known,
it is not at all clear which would be the best permutational isomer to attempt
a connection with, since the endpoints are so different. 
The distance between these structures was minimised with respect to permutation-inversion
isomers, centre of mass, and orientational coordinates.
However, this connection still presents a significant challenge, and
required 383 cycles of the Dijkstra-based algorithm to achieve a complete path
(Figure \ref{fig:C60}).

An enlarged view of
the pathway for the last two steps is also shown in Figure \ref{fig:C60}.
Both rearrangements correspond to the SW process mentioned above \cite{stonew86}.
The first step converts a patch containing seven, six, and two five-membered rings
into a patch with three six-membered rings and one five-membered ring.
The second step involves a more conventional process linking two patches
that both contain two six-membered and two five-membered rings.
The energy profile in this part of the pathway is consistent with the pattern
of high barriers previously discussed for the low-energy region of the
PES \cite{walshw98,walesmw98,KumedaW03}.
The present results suggest that further investigation of paths involving
non-fullerene C$_{60}$ isomers may be worthwhile, although we note once again
that this initial path may not be dynamically significant.

\subsection{GB1 Hairpin}

The GB1 hairpin consists of residues 41-56 from the C-terminal fragment
of the B1 domain of protein G. 
It forms a $\beta$-hairpin both in the complete protein \cite{gronenbornfeawwc91}, and for the
isolated fragment in solution \cite{blancors94}.
A number of previous experimental and theoretical studies have been conducted
for this system \cite{blancos95,munozthe97,munozhhe98,dinnerlk99,roccatanoadb99,man00,%
garcias01,zhoubg01,zagrovicsp01,%
zhoub02,bolhuis03,weidm03,KrivovK04},
including a DPS investigation \cite{EvansW04}.
Obtaining an initial path between unfolded and hairpin conformations in the
latter study was not an easy task, and the new algorithm speeded
up this part of the calculation by at least a couple of orders of magnitude.

\begin{figure}
\psfrag{V}[bl][bl]{$V$}
\psfrag{s}[bc][bc]{$s$}
\psfrag{-480}[cr][cr]{$-480$}
\psfrag{-490}[cr][cr]{$-490$}
\psfrag{-500}[cr][cr]{$-500$}
\psfrag{-510}[cr][cr]{$-510$}
\psfrag{-520}[cr][cr]{$-520$}
\psfrag{-530}[cr][cr]{$-530$}
\psfrag{-540}[cr][cr]{$-540$}
\psfrag{0}[tc][tc]{0}
\psfrag{500}[tc][tc]{500}
\psfrag{1000}[tc][tc]{1000}
\psfrag{1500}[tc][tc]{1500}
\psfrag{2000}[tc][tc]{2000}
\psfrag{2500}[tc][tc]{2500}
\psfrag{3000}[tc][tc]{3000}
\psfrag{3500}[tc][tc]{3500}
\psfrag{4000}[tc][tc]{4000}
\centerline{\includegraphics[width=1.25\textwidth]{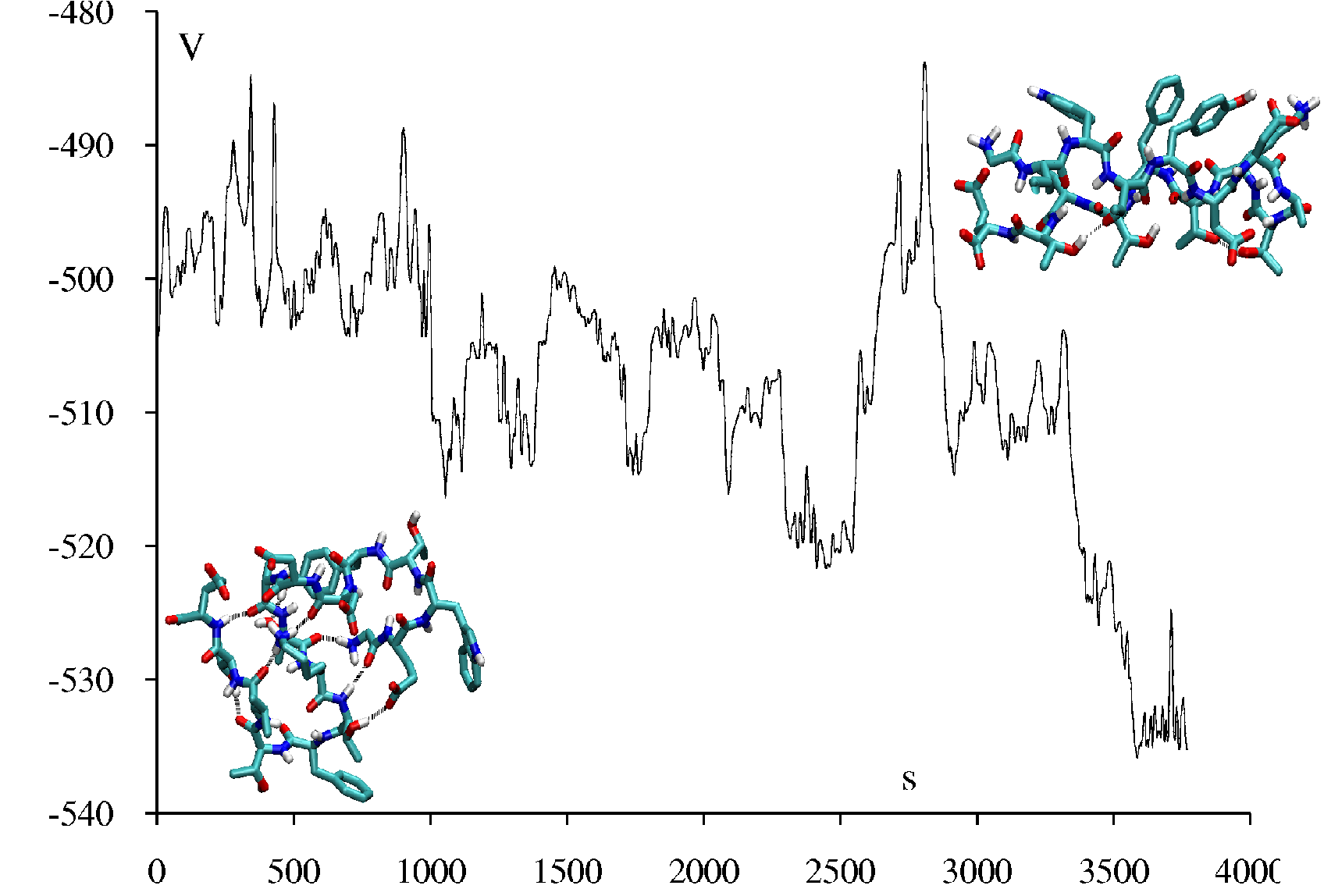}}
\ \vfill
\caption{}
\label{fig:GB1}
\end{figure}

The {\tt CHARMM} program\cite{brooksbossk83} has previously been interfaced to our
{\tt OPTIM} code \cite{download}, which includes a wide variety of algorithms for
locating stationary points and characterising pathways, and now includes
the Dijkstra-based connection approach.
As for the previous DPS study we employed the CHARMM19 force field with the
EEF1 implicit solvation model \cite{lazaridisk99}. 
The path illustrated in Figure \ref{fig:GB1} consists of 163 transition states,
and was found in 126 cycles of the Dijkstra-based algorithm.
It connects a partially collapsed non-native structure, with two turns, to
a conformation from the native hairpin ensemble.

\subsection{Tryptophan Zippers}

Tryptophan zippers are stable fast-folding $\beta$-hairpins designed by Cochran
{\it et al.}~\cite{cochranss01}, which have recently generated considerable 
interest.\cite{snowqdghp04,duzhg04} In the present work we have obtained native to
denatured state rearrangement pathways for five tryptophan zippers: trpzip $1$, trpzip $2$, trpzip $3$,
trpzip $3$-I and trpzip $4$. The notation is adopted from the work of Du
{\it et al.}~\cite{duzhg04}. All these peptides contain twelve residues, except for trpzip
$4$, which has sixteen. Tryptophan zippers $1$, $2$, $3$ and $3$-I
differ only in the sequence of the turn. Experimental measurements of characteristic folding
times for these peptides have shed some light on the significance of the turn sequence in 
determining the stability and
folding kinetics of peptides with the $\beta$-hairpin structural motif.\cite{duzhg04}

To model these molecules we used a modified CHARMM19 force field~\cite{brooksbossk83}, 
with symmetrised ASN, GLN and TYR
dihedral angle and CTER improper dihedral angle terms, to ensure that rotamers of these residues 
have the same energies and geometries.\cite{rotamersurl} Another small modification
concerned the addition of a non-standard amino acid, D-proline, which was needed to model trpzip $3$.
The implicit solvent model EEF1 was used to account for solvation,\cite{lazaridisk99}
with a small change to the original implementation to
eliminate discontinuities.\cite{bloomthesis}

We have used the Dijkstra-based connection algorithm to obtain folding pathways for 
all five trpzip peptides. 
In each case the first endpoint was chosen to be the native state structure, which, for
$1$, $2$ and $4$ trpzips, was taken from the Protein Data Bank (PDB).\cite{pdb}
There are no NMR
structures available for $3$ and $3$-I, so for these
peptides the first endpoint was chosen to be the putative global minimum obtained
using the basin-hopping method.\cite{lis87,walesd97,waless99}
The second endpoint was chosen to be an extended structure, which was obtained by
simply minimising the energy of 
a conformation with all the backbone dihedral angles set to 180 degrees.
All the stationary points (including these obtained during the connection procedure) 
were tightly converged to reduce the root-mean-squared force below 
$10^{-10}$\,kcal\,mol$^{-1}$\,\AA$^{-1}$.

\begin{figure}
\centerline{\includegraphics[height=0.95\textheight]{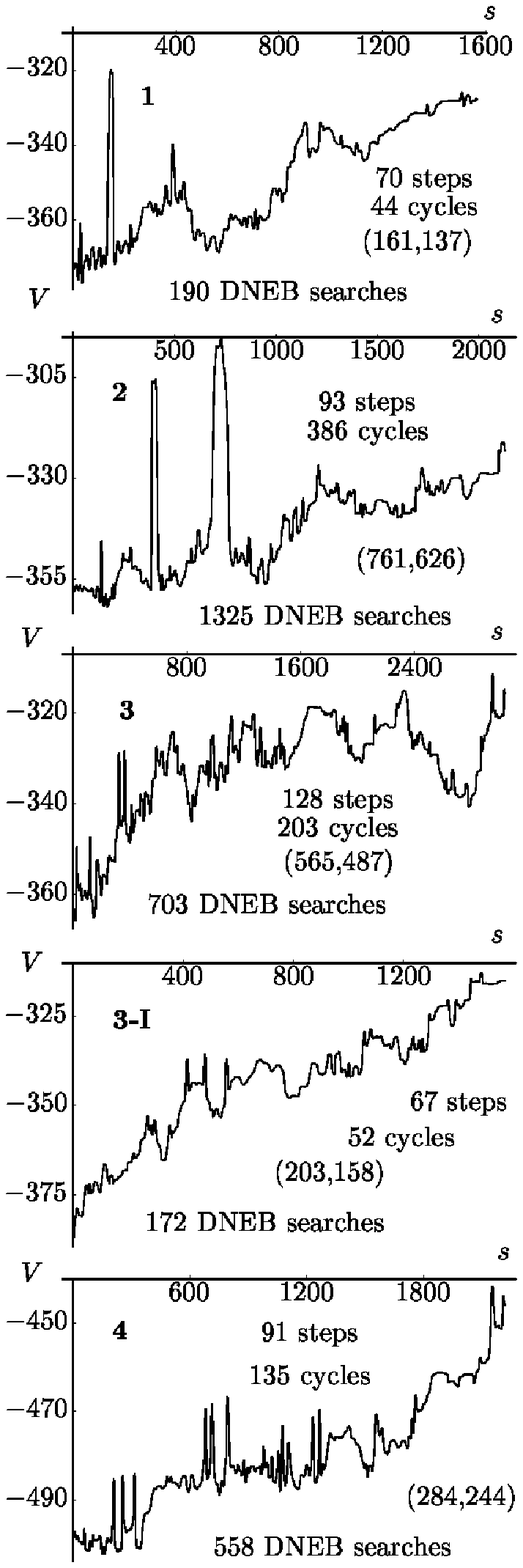}}
\ \vfill
\caption{}
\label{fig:trpzipEofS}
\end{figure}

Each of the five trpzip pathway searches was
conducted on a single Xeon 3.0\,GHz CPU and required less than 24 hours of CPU time.
The timings could certainly be improved by optimising the various parameters 
employed throughout the searches. 
However, it is more important that the connections actually
succeed in a reasonably short time. 
It only requires one complete path to seed a DPS run, and we expect the DPS procedure
to reduce the length of the initial path by a least a factor of two in sampling the
largest contributions to the effective two-state rate constants.
The results of all the trpzip calculations are shown in Figure \ref{fig:trpzipEofS}.

\subsection{Villin Headpiece Subdomain}

The villin headpiece subdomain is the thermostable 35-residue C-terminal section 
of the headpiece domain of chicken villin protein.  
The sequence used for
the NMR structure determination~\cite{mcknightmk97} included an additional methionine 
residue at the N-terminus from the expression system; thus, we are considering the 36-residue
entity here, PDB code 1VII.  
The structure consists of a bundle of three short helices and a closely-packed hydrophobic core.
The three helices are numbered from the N-terminus to the C-terminus.  A turn connects 
helices 2 and 3, and helices 1 and 2 are joined by a loop.  

Its small size and fast folding (a folding time on the order of tens
of microseconds~\cite{kubelkaeh03,wangtsvmr03}) make the villin headpiece subdomain an
attractive target for computational studies, 
including the $1\,\mu{\rm s}$ explicit-water MD simulation of Duan 
and Kollman.~\cite{duank98,duanwk98,leedk00,sullivank02,shenf02,wenhkl04,
jangksp03,fernandezscsbf03,mukherjeeb03,mukherjeeb04,islamkw02,pandebceklrsssz03,zagrovicssp02}
Our DPS study of the villin headpiece subdomain employed the UNRES (united-residue) force field and 
model~\cite{liwoopwrs97,liwopwros97,liwokcgowrps98}, 
in which two interaction sites are assigned to each residue: 
one representing the main chain peptide group and one 
representing the side chain.

\begin{figure}
\psfrag{V}[bl][bl]{$V$}
\psfrag{s}[bc][bc]{$s$}
\psfrag{-100}[cr][cr]{$-100$}
\psfrag{-105}[cr][cr]{$-105$}
\psfrag{-110}[cr][cr]{$-110$}
\psfrag{-115}[cr][cr]{$-115$}
\psfrag{-120}[cr][cr]{$-120$}
\psfrag{-125}[cr][cr]{$-125$}
\psfrag{-130}[cr][cr]{$-130$}
\psfrag{0}[tc][tc]{0}
\psfrag{N}[cl][cl]{N}
\psfrag{C}[cl][cl]{C}
\psfrag{1}[br][br]{1}
\psfrag{2}[tl][tl]{2}
\psfrag{3}[cc][cc]{3}
\psfrag{500}[tc][tc]{500}
\psfrag{100}[tc][tc]{100}
\psfrag{200}[tc][tc]{200}
\psfrag{300}[tc][tc]{300}
\psfrag{400}[tc][tc]{400}
\centerline{\includegraphics[width=1.25\textwidth]{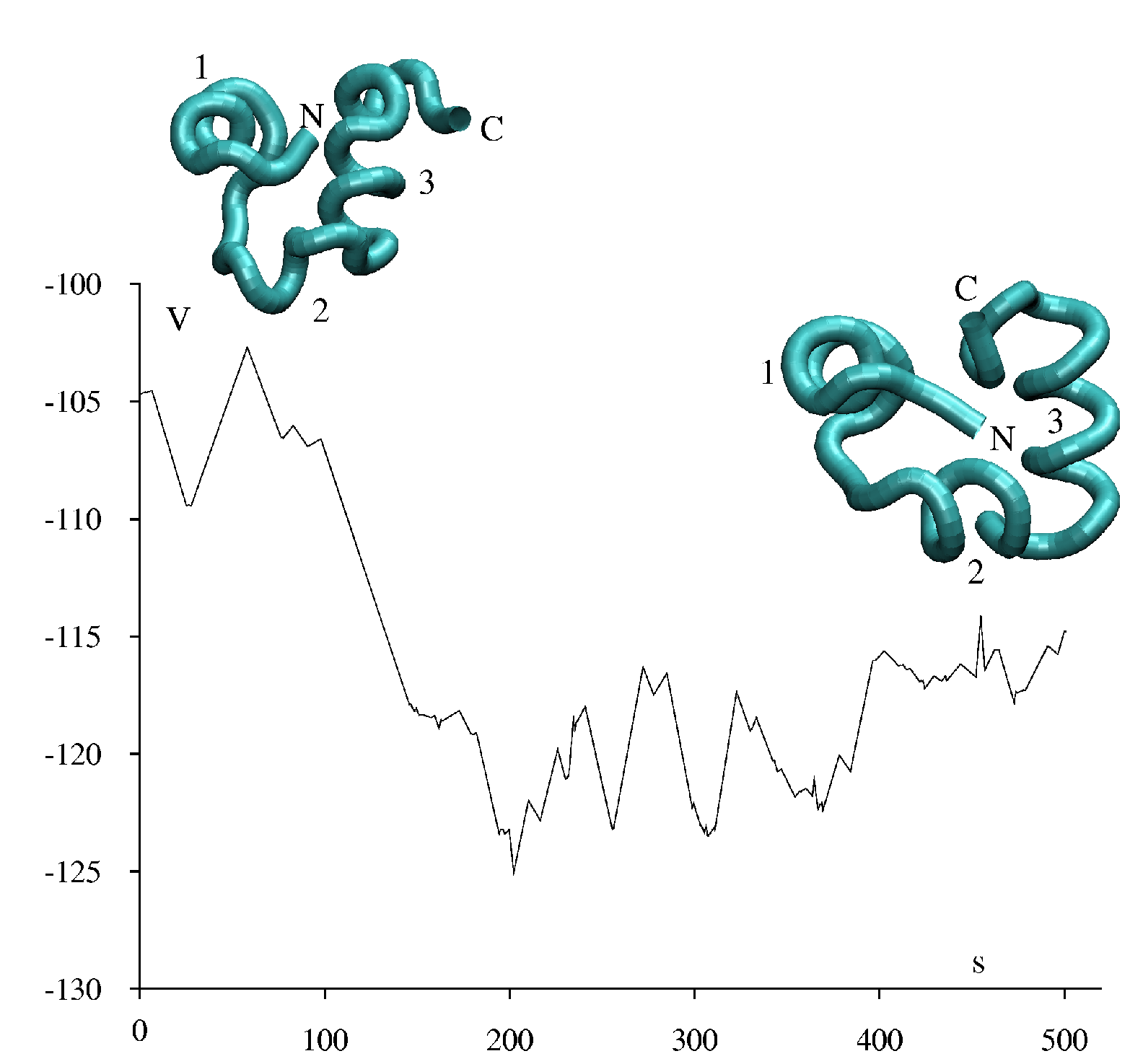}}
\ \vfill
\caption{}
\label{fig:villin}
\end{figure}

As for the GB1 hairpin, the construction of an initial path with the original connection algorithm proved 
difficult, and a considerable increase in both the efficiency and the success rate was attained with the 
Dijkstra-based strategy.  
The example 
path illustrated in Figure \ref{fig:villin} contains 62 transition states and required 142 cycles of the 
Dijkstra-based algorithm.  In this rearrangement, which connects a partially folded minimum to the locally 
minimised PDB structure, helices 2 and 3 are completed, the C-terminal residues pack correctly between 
helices 1 and 3, and the three helices adopt their native relative orientations. 

\section{Conclusions}

Discrete path sampling calculations of effective two-state rate constants require an
initial path consisting of local minima and the transition states that
connect them \cite{Wales02,Wales03,Wales04,EvansW03b,CalvoSW03,EvansW04}.
For complex rearrangements the number of elementary steps involved may be
rather large, and new methods for constructing an initial path are needed. 
This path does not need to be the shortest, or the fastest, but it does need
to be fully connected.
In the present work we have described a connection procedure based upon Dijkstra's
shortest path algorithm, which enables us to select the most promising paths
that include missing connections for subsequent double-ended searches.
We have found that this approach, 
which is now implemented within the {\tt OPTIM} package \cite{download},  
enables initial paths containing more than a
hundred steps to be calculated automatically for a variety of systems.
Some typical results have been presented for buckminsterfullerene, trpzip peptides,
the GB1 hairpin, and the villin headpiece subdomain.
These paths will be employed to seed future discrete path sampling calculations.

\section{Acknowledgements}

We gratefully acknowledge Dr David Evans for his Dijkstra-based utilities that
locate shortest paths in connected databases of minima and transition states.
S.A.T.~is a Cambridge Commonwealth Trust/Cambridge Overseas Trust scholar. Most
of the tryptophan zippers calculations were performed using computational facilities 
funded by the Isaac Newton Trust. J.M.C.~is grateful to the EPSRC for financial support.

\bibliographystyle{thesis}
\bibliography{Dijkstra}
\clearpage

\section*{Figure Captions}
\begin{enumerate}
\item 
Top: connected path between a high-energy C$_{60}$ cluster and buckminsterfullerene.
The path contains 82 transition states, and required 383 cycles of the Dijkstra-based
connection algorithm, including 1620 DNEB searches.
$V$ is the potential energy in hartree and $s$ is the integrated path length in bohr.
Bottom: enlargement of the above plot for the last two SW rearrangements. 
The atoms mainly involved in these two steps are shaded, and both local 
minima and transition state structures are indicated at appropriate points along the path.
\item
Energy profiles for native to denatured state rearrangements of tryptophan
zippers found by the Dijksta-based connection algorithm. For each profile the number of steps in
the pathway, the number of connection algorithm cycles, the total number of DNEB
searches and the total number of stationary points in the database (recorded upon termination of the
algorithm) are shown. The total number of stationary points is presented in the form $(n,m)$, where
$n$ is the number of minima and $m$ is the number of transition states.
The potential energy, $V$, is given in the units of kcal/mol, and the integrated path
length, $s$, is given in the units of \AA.
\item 
Connected path between non-native and $\beta$-hairpin conformations for the
GB1 peptide. The path contains 163 transition states, and required 126 cycles of the Dijkstra-based
connection algorithm, including 1610 DNEB searches.
$V$ is the potential energy in kcal/mol and $s$ is the integrated path length in \AA.
\item 
Connected path between non-native and native conformations for the
villin headpiece subdomain. 
The two endpoint minima are shown with the N-terminal helix (helix 1) on the left.
The other two helices and the C-terminus are also labelled.
The path contains 62 transition states, and required 142 cycles of the Dijkstra-based
connection algorithm, including 294 DNEB searches.
$V$ is the potential energy in kcal/mol and $s$ is the minimised Euclidean distance between 
consecutive stationary points in \AA. For this system we simply join the stationary
points with straight lines.
\end{enumerate}

\clearpage

\pagestyle{empty}
\clearpage

\pagestyle{empty}
\clearpage

\clearpage

\clearpage

\end{document}